Magnetic anomalies in a spin-chain compound, $Sr_3CuRhO_6$: Griffiths-phase-like behavior of magnetic susceptibility

E.V. Sampathkumaran, Niharika Mohapatra, Sudhindra Rayaprol and Kartik K Iyer
Tata Institute of Fundamental Research, Homi Bhabha Road, Colaba, Mumbai-400005,
India

We report the results of ac and dc magnetic susceptibility ($\chi$), heat-capacity (C), isothermal magnetization and isothermal remanent magnetization measurements on the compound, $Sr_3CuRhO_6$, crystallizing in a $K_4CdCl_6$-derived monoclinic structure. The magnetization data reveal distinct magnetic anomalies near 6 and 12 K with decreasing temperature (T). While the transition below ($T_o$=) 6K appears to be of a spin-glass-type as inferred from all the data, the one at 12 K is not typical of bulk ferromagnetism in contrast to an earlier proposal. In the range 6 to 12 K, the dc $\chi$ obeys the form, $\chi^{-1} \alpha (T-T_o)^{1-\lambda}$ (0<$\lambda$<1) for very low dc fields and the values of $\chi$ decrease gradually with the application of higher magnetic fields, mimicking the behavior Griffiths-phases.

PACS numbers: 75.40.Cx, 75.30.Kz, 75.50.-y, 71.27.+a

## I. Introduction

The spin-chain compounds of the type, $(Sr,Ca)_3MXO_6$ (M, X= transition metal ions), with M-X magnetic chains running along *c*-axis separated by Sr (Ca) ions, derived from $K_4CdCl_6$-type rhombohedral structure (space group $R\bar{3}c$) have attracted considerable interest in recent years [1-3]. $MO_6$ (distorted) trigonal prism and $XO_6$ (distorted) octahedra share a face. In general, intrachain interaction is ferromagnetic, whereas interchain interaction is antiferromagnetic. The magnetic chains are placed in a triangular fashion in the basal plane and hence often geometrically frustrated magnetism, particularly with unusual features, are observed among this class compounds. The Cu containing compounds (at the M site) are especially interesting, as these undergo monoclinic distortion (space group $C2/c$) due to Jahn-Teller effect, as a result of which Cu ions are displaced alternately to the right and to the left with respect to the X-ion chain in a zig-zag fashion, thereby tending towards square-planar coordination (see Ref. 1 for details). In this article, we present the results of our investigations on $Sr_3CuRhO_6$, particularly considering that a previous preliminary report based on *dc* magnetization studies [1] claimed that this compound is ferromagnetic below ($T_C=$) 12 K.

## II. Experimental details

The sample in the polycrystalline form was prepared via a solid state route. Requisite amounts of $SrCO_3$, CuO and Rh powder (purity of which is more than 99.99%) were intimately mixed and calcined at 800 C for 1 day and 900 C for 1 day with an intermediate grinding. The calcined powder was then ground well, pelletized and sintered at 1150 C over a period of 15 days with few intermediate grindings. The sample thus obtained was characterized by x-ray diffraction (Cu $K_\alpha$) and found to be single phase within the limits of x-ray diffraction. Further composition homogeneity was verified by scanning electron microscope. The lattice parameters ($a=$ 9.225(3) Å, $b=$ 9.688(3) Å, $c=$ 6.693(3) Å, and $\beta=$ 92.46º) are found to be in excellent agreement with those reported in Ref. 1. The magnetic susceptibility ($\chi$) measurements in the temperature interval 2-300 K in the presence of several *dc* magnetic fields (H=2, 20, 50 and100 Oe, 1, 5 and 10 kOe) and isothermal magnetization (M) measurements at several temperatures were performed with a commercial superconducting quantum interference device (SQUID) (Quantum Design, USA). Additional measurements were performed with a commercial vibrating sample magnetometer (VSM) (Oxford Instruments, UK) to confirm the main findings. *Ac* $\chi$ data (2 – 30 K) were also obtained with the same SQUID magnetometer ($H_{ac}=$ 1 Oe) with few frequencies. Heat-capacity (C) data were collected in the range 1.8-20 K employing a commercial Physical Property Measurements System (PPMS) (Quantum Design, USA) by relaxation method.

## III. Results and discussion

*We first address (a) whether both Cu and Rh contain magnetic moment and (b) whether there is a long-range magnetic transition temperature.*

For this purpose, we first show in figure 1 the results of $\chi$ measurements measured in the presence of 100 Oe and 5 kOe down to 1.8 K. We have taken the data for the zero-field-cooled (ZFC) and field-cooled (FC) conditions of the specimens. The $\chi$ obeys Curie-Weiss behavior above 150 K (figure 1a) and the value of the paramagnetic Curie temperature ($\theta_p$) obtained from the linear region is close to –120 K with the negative sign indicating the existence of antiferromagnetic correlations. As the temperature is lowered, there is a deviation from the high temperature linearity in $\chi^{-1}(T)$ plot, similar to that reported for $Sr_3CuIrO_6$ [Ref. 4] in which case it was attributed to gradual dominance of clustering effects. The effective magnetic moment at high temperatures (>150 K) is found to be 2.60 $\mu_B$, supporting the findings of Ref. 1, consistent with spin-only moment for $Rh^{4+}$ (low-spin at the octahedral site, $t_{2g}^5$) and $Cu^{2+}$ in a square planar coordination. Thus, there is a magnetic moment on both Rh and Cu. Since the inter-chain separation is nearly twice of Cu-Rh distance, it is possible that this is a quasi-1-dimensional magnetic compound. With respect to the low temperature, the ZFC-FC curves overlap for H= 5 kOe with a monotonic increase of magnitude with decreasing temperature, while the curves tend to bifurcate below about 6 K for H= 100 Oe (figure 1b). However, there is a peak for ZFC curve for H= 100 Oe near 6 K indicating the existence of a magnetic transition at this temperature. Further comparison of low temperature $\chi(T)$ shape for these fields is quite revealing. For H= 100 Oe, there is a sudden upturn below 12 K, and on the basis of a similar observation for H= 500 Oe, it was claimed in Ref. 1 that long range ferromagnetic order sets in at this temperature. However, this variation of $\chi$ is rather smooth around this temperature for H= 5 kOe. These observations imply that there is a considerable ambiguity in assigning magnetic ordering temperature ($T_o$).

In order to throw more light on the magnetic ordering behavior, we have performed heat-capacity measurements. The heat-capacity results are shown in figures 1c and 1d. C monotonically decreases with decreasing temperature down to 1.8 K without any evidence for a prominent $\lambda$-anomaly typical of bulk magnetic ordering, though an extremely weak peak could be seen near 11 K if the data is plotted in the form of C/T *versus* T. In fact, C(T) curves for H= 10 and 50 kOe overlap with that of H= 0 down to at least 6 K. We believe that the absence of any prominent feature near 12 K in C(T) can be taken as an evidence for the absence of bulk magnetic order at this temperature; however, it is not clear whether there is a role of short-range magnetic correlations preexisting at higher temperatures in quasi-low-dimensional systems gradually removing entropy thereby washing away a prominent $\lambda$-anomaly as proposed by a recent theory [5]. A careful look at the plot of C/T (figure 1c) for H= 0 shows an upturn below 6 K. This upturn could be suppressed by the application of a magnetic field, say, 50 kOe (see figure 1c), which could be taken as indication for the onset of a bulk magnetic transition at this temperature. Note that this is nearly the same temperature at which there is a peak in ZFC-$\chi(T)$ for H= 100 Oe. Combined with the fact that there is a bifurcation of ZFC-FC $\chi(T)$ curves, the above weak C feature suggests the onset of spin-glass freezing near 6 K. Finally, C/T varies linearly in the range 6 to 20 K (see figure 1d) and the Debye temperature derived from this linear region turns out to be close to 330 K. Despite the fact that this compound is an insulator, the linear coefficient ($\gamma$) of C extrapolated to T = 0 is apparently large (nearly 100 mJ/mol $K^2$); the significance of this finding is not clear, though such a feature is also typical of spin glasses. We do not have adequate knowledge at present on the details of various contributions to entropy below 12 K in this system to understand this better.

We now present M(H) data to support of the existence of spin-glass anomalies below 6 K. In support of the existence of a ferromagnetic component below 6 K, we find that M(H) is hysteretic (see 1.8 K data in Fig. 2a for a typical loop). However, the high field behavior of M(H) (see figure 2b), following a sharp rise at low fields, exhibits a monotonic increase with H till the measured range of 70 kOe without any evidence for a saturation as though there is an antiferromagnetic component as well. In addition, the absolute values of M at high magnetic fields is too small (far below 1 $\mu_B$) that the compound can not be classified as a ferromagnet (not even as an itinerant ferromagnet, considering that these materials are insulators). These magnetization features are typical of spin-glasses (below 6 K in this case). To support this conclusion, the isothermal remnant magnetization ($M_{IRM}$) were performed. For this purpose, we took M data as a function of time ($t$) in zero-field after magnetizing the sample in 5 kOe for 5 minutes for the ZFC condition of the specimen; $M_{IRM}$ is distinctly non-zero for $t=0$ (0.034 and 0.023 $\mu_B$ per formula unit respectively for 2 and 5 K) and it exhibits a slow decay varying logarithmically with $t$ (see figure 2c for a typical behavior). Similar experiments in the paramagnetic state, say at 20 K, resulted in nearly zero value for $M_{IRM}$ at $t=0$. Turning to the magnetic behavior above 6 K, the hysteresis loop collapses when temperature is raised beyond 6 K as shown typically for 7.5 K in figure 2a, thereby establishing that the magnetic states above and below 6 K are different. $M_{IRM}$ behavior at 7.5 K is even found to be qualitatively different from those below 6 K and offers further convincing evidence for this conclusion: At $t=0$, the value is found to be negligibly small (0.001 $\mu_B$ per formula unit) without any decay (though we observed a gradual small increase, say by about 1% in 1 hour, the implications of which is not clear to us at present).

*We now focus on the behavior of 12K- transition in the low field $\chi(T)$ in figure 1b.*

We have performed $\chi(T)$ measurements at several dc fields in the range 20 Oe to 10 kOe above 5 K, which are shown in figure 3 in the form of $\chi^{-1}$ versus T below 30 K. We find that the plots above nearly 15 K tend to overlap for all H (and hence not shown above 30 K in figure 3). However, for low fields, say below 50 Oe, there is a dramatic fall of $\chi^{-1}$ below about 12 K followed by a slower variation in the range 6 to 10 K. As the magnetic field is increased, the absolute values of $\chi$ tend to decrease and the nature of the fall in the inverse $\chi$ in figure 3 in the range 10 to 12 K gets less sharp. Viewed together with the shape of M(H) curve at 7 K (see figure 2b), these observations imply that (ferro)magnetic components are present in the material, the magnetization of which can be gradually saturated with increasing fields. The features reveal as though the linear response of the paramagnetic component gets gradually highlighted with increasing H. It may be recalled that the observed H-dependence of $\chi^{-1}(T)$ plots, seen in figure 3, is the fingerprint [6,7] of Griffiths phase.

At this moment, it is worth recalling that the concept of 'Griffiths phase' was proposed few decades ago [8,9] following an original idea of Griffiths [10] to describe a situation in which the crystallographic disorder suppresses the long-range ferromagnetic ordering temperature, resulting in an intermediate temperature region in which ferromagnetic clusters are dispersed in a paramagnetic matrix. The intra-cluster magnetic ordering temperature above experimentally observable bulk $T_C$ varies depending on the size of the cluster and therefore there is a spectrum of $T_C$ below the characteristic Griffiths temperature, $T_G$. Though the existence of such a magnetic phase due to quenched disorder has been discussed in connection with giant magnetoresistance behavior of manganites [6, 11-13], non-Fermi liquid behavior in f-electron systems [14,15], and dilute magnetic semiconductors [16] as well as for a Tb-based intermetallic compound [7], it is believed that the experimental evidence in favor of the existence of Griffiths phase is still inadequate [17].

In view of this situation, we wonder whether the features seen in figure 3 in the range 6-12 K signals Griffiths phase region. Negligibly small values of $M_{IRM}$ at $t=0$ mentioned earlier implies that globally there is no spontaneous magnetization as expected for Griffiths phase. It is known [6-9, 14] that the $\chi$ in the Griffiths phase is characterized by an exponent lower than unity, i.e., $\chi^{-1}(T) \alpha (T-T_o)^{1-\lambda}$ ($0<\lambda<1$). Therefore, we have fitted logarithm of $\chi^{-1}$ in the range 6 to 10 K, say for H= 20 Oe data, and *the value of λ turns to be 0.8 (see figure 3, inset), in agreement with theoretical predictions* [14], whereas in the range 12-20 K, it is close to zero. It will be interesting to probe by small angle neutron scattering or muon spin rotation measurements to explore this scenario. In this connection, as mentioned earlier, C/T varies rather linearly with $T^2$ (see figure 1d), typical of phonon contribution in the range 6 – 12 K, in contrast to predicted forms $(T-T_o)^{1-\lambda}$ ($0<\lambda<1$) for Griffiths phase scenario among non-Fermi liquid systems [14,15].

We have performed *ac* $\chi$ measurements in order to compare the response in the two relevant temperature regions. The results are shown in figure 4. There is a well-defined peak in both real ($\chi'$) and imaginary ($\chi''$) parts at 6 K – at the same temperature where low-field ZFC-FC curves of dc $\chi(T)$ bifurcate and bulk spin-glass freezing is proposed to occur in the above discussions. However the peaks at 6 K are not sufficiently sharp enough to resolve any frequency dependence [18] characterizing canonical spin glasses, particularly noting that the data is noisy at higher frequencies (say 1339 Hz). (For the same reason, we have not shown the data for the imaginary part at higher frequencies). However, there is an additional broad peak in both the components in the range 6 to 10 K following a sharp upturn near 12 K - in those temperature intervals where there is no evidence for long range magnetic ordering of any kind. It is this width of this peak that demands an explanation which is different from 'conventional spin glass'. We therefore associate such a feature to domain dynamics of weak ferromagnetism characterizing this phase. All these peaks vanish for an application of a magnetic field of 10 kOe, typical of disordered magnetism.

### IV. Conclusion

To conclude, we have reported *ac* and *dc* magnetization and heat capacity behavior of $Sr_3CuRhO_6$. It appears that this compound is not a bulk ferromagnet below 12 K and that there is another magnetic transition of a spin glass type below 6 K. The magnetism in the intermediate temperature range (6-12 K) appears to be of an inhomogeneous type with a weak ferromagnetic component, characterized by a magnetic susceptibility behavior mimicking Griffiths-phase. We call for further microscopic studies to understand the exact nature of the magnetic phase in this intermediate temperature range. Finally, it is important to mention that we have studied few other materials of this family, $Sr_3NiRhO_6$ and $Sr_3NiPtO_6$, in which there is no structural distortion due to the absence of Jahn-Teller ion, and we do not find the $\chi(H,T)$ anomaly seen in this Cu-based material. This finding indicates that Jahn-Teller effect plays a crucial role in inhomogeneous magnetism.

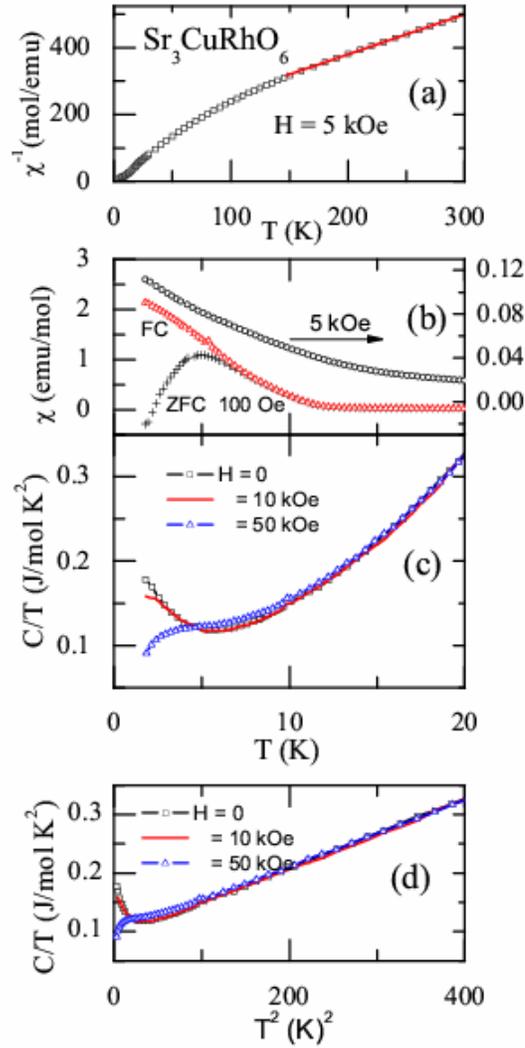

Figure 1:
(color online) *(a)* Inverse magnetic susceptibility ($\chi$) as a function of temperature (T) taken in the presence of a *dc* magnetic field (H) of 5 kOe, and *(b)* $\chi$(T) obtained in the presence of H= 100 Oe for the zero-field-cooled (ZFC) and field-cooled (FC) conditions of the specimen, $Sr_3CuRhO_6$. A straight line is drawn through the high temperature linear region in *(a)*. In *(c)* and *(d)*, heat-capacity (C) as a function of temperature, also taken in the presence of magnetic fields, is shown in various ways.

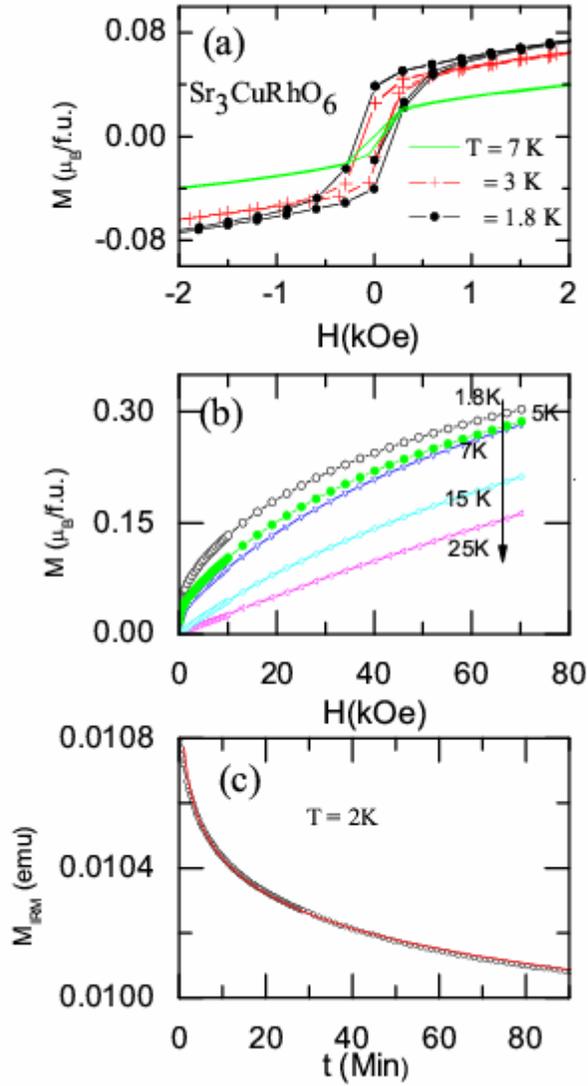

Figure 2:
(color online) *(a)* Magnetic hysteresis loops at 1.8 and 7.5 K, *(b)* isothermal magnetization at selected temperatures, and *(c)* isothermal remnant magnetization curve for T= 2 K obtained as mentioned in the text, for $Sr_3CuRhO_6$. While the continuous lines through the data points serve as guides to the eyes, in *(c)*, the line represents logarithmic fitting to the form 1- 0.011log *t*.

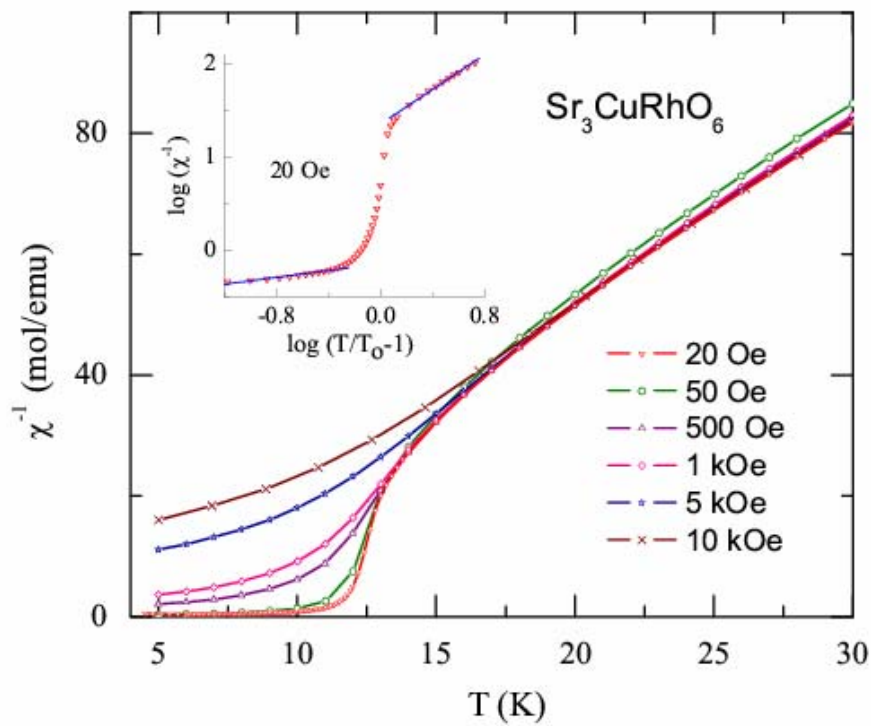

Figure 3:
(color online) Inverse magnetic susceptibility as a function of temperature for $Sr_3CuRhO_6$ obtained with a SQUID magnetometer in the presence of several *dc* magnetic fields. Continuous lines drawn are guides to the eyes. In the inset, the data for H= 20 Oe are plotted as $\log(\chi^{-1})$ versus $\log(T/T_o-1)$, assuming that that $T_o$= 6 K; the continuous lines above and below 12 K are due to fitting.

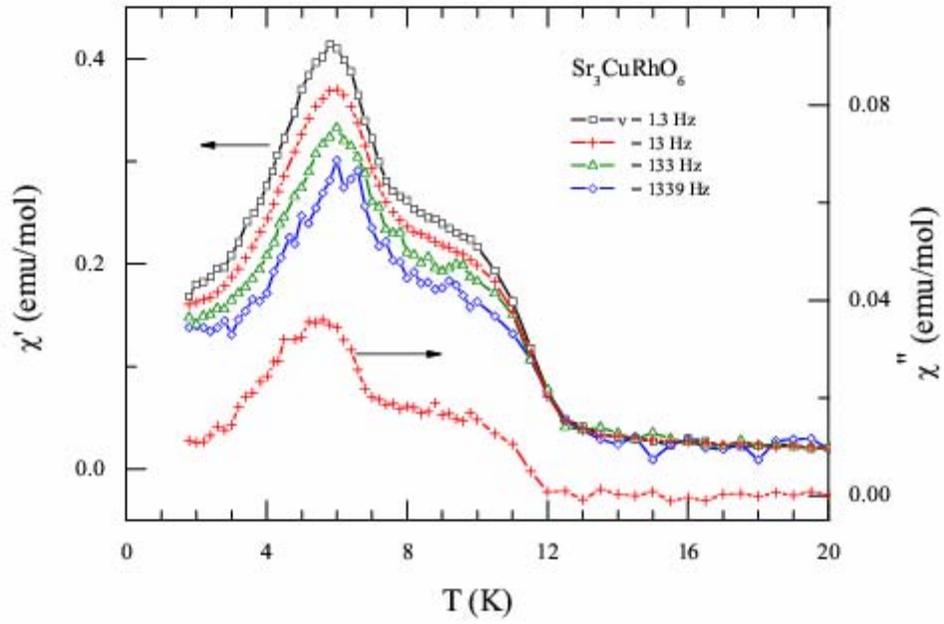

Figure 4:
(color online) Real ($\chi'$) and imaginary ($\chi''$) parts of *ac* magnetic susceptibility as a function of temperature for $Sr_3CuRhO_6$ below 20 K.